\documentclass[12pt]{article}

\usepackage{amsmath,amssymb}

\newcommand{\R}{{\mathbb{R}}}
\newcommand{\Z}{{\mathbb{Z}}}

\newcommand{\T}{{\mathbb{T}}}
\newcommand{\M}{{\mathbb{M}}}

\newcommand{\ba}{\begin{array}}
\newcommand{\ea}{\end{array}}

\newcommand{\bp}{\begin{pmatrix}}
\newcommand{\ep}{\end{pmatrix}}

\newcommand{\bps}{\begin{smallmatrix}}
\newcommand{\eps}{\end{smallmatrix}}

\newcommand{\f}{\frac}

\newcommand{\no}{\nonumber}

\def \J{{\bf J}}
\def \cJ{{\cal J}}

\def \cA{{\cal A}}
\def \cB{{\cal B}}
\def \cC{{\cal C}}
\def \cD{{\cal D}}

\def \cAb{{\bar {\cal A}}}
\def \cBb{{\bar {\cal B}}}
\def \cCb{{\bar {\cal C}}}
\def \cDb{{\bar {\cal D}}}

\def \Xd{{\dot X}}
\def \Xa{{\acute X}}

\def \raw{\rightarrow}

\def \rank{\mathrm{rank}}

\def \half{\frac{1}{2}}
\def \ov#1{\frac{1}{#1}}

\def \cA{{\cal A}}

\def \fpart#1#2{\frac{\partial #1}{\partial #2}}

\def \fdel#1#2{\frac{\delta {#1}}{\delta {#2}}}

\def \l{{\frak l}}

\def \cM{{\cal M}}

\def \l({\left(}
\def \r){\right)}

\def \0{{\bf 0}}
\def \1{{\bf 1}} 

\def \Mat{\mathit{Mat}}

\begin{document}

\begin{titlepage}
\thispagestyle{empty}
\begin{flushleft}
\hfill hep-th/0105056\\
UTMS 2001-12\hfill May, 2001 \\
\end{flushleft}

\vskip 1.5 cm

\begin{center}
\noindent{\Large \textbf{T-Duality Group for Open String Theory}}\\
\noindent{
 }\\
\renewcommand{\thefootnote}{\fnsymbol{footnote}}

\vskip 2cm
{\large 
Hiroshige Kajiura 
\footnote{e-mail address: kuzzy@ms.u-tokyo.ac.jp}\\

\noindent{ \bigskip }\\

\it
Graduate School of Mathematical Sciences, University of Tokyo \\
Komaba 3-8-1, Meguro-ku, Tokyo 153-8914, Japan\\
\noindent{\smallskip  }\\
}

\bigskip
\end{center}
\begin{abstract}
We study T-duality for open strings on tori $\T^d$. The general boundary 
conditions for the open strings are constructed, and it is shown that 
T-duality group, which preserves
 the mass spectrum of closed strings, preserves also the mass spectrum
 of the open strings. The open strings are transformed to 
those with different boundary conditions by T-duality. We also discuss the 
T-duality for D-brane mass spectrum, and show that the D-branes and 
the open strings with both ends on them are transformed together 
consistently.

\end{abstract}
\vfill

\end{titlepage}
\vfill
\setcounter{footnote}{0}
\renewcommand{\thefootnote}{\arabic{footnote}}
\newpage

\section{Introduction}

Noncommutative(NC) tori arise when compactifying matrix models 
on tori\cite{CDS}. 
When fixing the noncommutativity $\Theta$ of the base torus, 
there are still ambiguities for the choice of 
(NC) gauge theories on the NC torus, 
and the algebras of these gauge theories are Morita 
equivalent to each other (and Morita equivalent 
to the base NC torus). 
Morita equivalent two-tori are generated by 
$SL(2,\Z) (\times SL(2,\Z))$\cite{CDS}, and it is proven that, 
in general $d$ where $d$ is the dimension of the torus, 
Morita equivalent NC $\T^d$ is generated by $SO(d,d |\Z)$
\cite{S} 
and conversely NC $\T^d$ generated by $SO(d,d |\Z)$ is 
Morita equivalent\cite{RS,R} (for irrational NC parameter 
$\Theta^{ij}$). 
The Morita equivalence is realized geometrically in terms of 
quantum twisted bundles for two-tori\cite{Ho,MZ}, and further 
investigated for higher dimensional tori\cite{BMZ}. 

Originally, 
$SO(d,d |\Z)$ is known as the T-duality group, which acts on the 
closed string background $g, B$ so as to 
preserve the closed string mass spectrum\cite{GPR}. 
For this reason, this correspondence between Morita equivalence and 
$SO(d,d|\Z)$ is often said that 
`Morita equivalence is equivalent to T-duality'. 
However physically, the relation between this $SO(d,d |\Z)$ group 
(acting on NC parameter $\Theta$) and T-duality (acting on 
closed string background $g, B$) seems somehow not clear.  

On the other hand, in \cite{SW}, it is argued that in flat background 
$g, B$ (on $\R^d$  or $\T^d$), the field theory can be described by 
either commutative or noncommutative representation. 
The commutative description is natural for closed string theory, 
and noncommutative one is natural for open string theory 
(by the argument 
in \cite{SW}, it is natural when we construct the field theory 
from the OPE for the boundary of the open string). 
The transformation between the variables 
in the commutative picture $(g_s, g, B)$ and those in the NC 
picture $(G_s, G, \Theta)$ is also given
\footnote{Here $g_s$ are the closed string coupling constant, 
$G_s$ are the open string coupling constant, $G$ is the open string metric, 
and $\Theta$ is the noncommutativity of the space.}. 
Now on $\T^d$, T-duality group acts on commutative field theory side 
and Morita equivalence ($SO(d,d |\Z)$) is on NC side. 
Moreover for instance in \cite{SW} (for $|B/g| \raw\infty$ limit), 
the $SO(d,d|\Z)$ T-duality transformation 
on the NC side is defined 
by the action of the T-duality group on $(g_s, g, B)$ 
and using this transformation between $(g_s, g, B)$ and $(G_s, G, \Theta)$. 
By this definition, 
the transformation between $(g_s, g, B)$ and $(G_s, G, \Theta)$ is 
equivariant with the T-duality action on both sides
\footnote{This compatibility is true also for finite $|B/g|$ 
with a little modification\cite{PS}. }, 
and for any commutative theory generated by T-duality, 
there is one equivalent NC theory. 
This T-duality on NC side coincides the Morita equivalence 
for the action on $\Theta$. 

However, NC theory is the theory on the open string picture. 
Therefore instead of the above indirect realization of Morita equivalence 
or $SO(d,d|\Z)$ T-duality on NC space, 
we would like to realize it 
directly from open string physics, 
like that the realization of the noncommutativity 
on tori was studied from open strings \cite{DH,KO,CK}
\footnote{In ref. \cite{SW}, an interpretation on Morita equivalence 
from open strings is given in $|B/g|\raw \infty$ limit. We take a different 
approach and the results in the present paper are valid without taking the 
$|B/g|\raw \infty$ limit. }
. 

This is the motivation of the present paper. Towards this propose, 
in this paper, the general $(S)O(d,d|\Z)$ `T-duality covariant' forms 
of the open string modes and the T-duality invariant form of the 
open string mass spectrum are given explicitly. 
These arguments include the T-duality 
transformation for the boundary conditions of open strings. 
The T-duality transformation for the boundary conditions is considered 
from different angles : 
from the point of view of canonical transformations\cite{J}, 
by utilizing $O(d,d)$ symmetry
\footnote{The way used for closed string theory in \cite{MSch} is applied. } 
which is inherent in the form of the open string Hamiltonian\cite{Ma}, 
and by the boundary state formalism\cite{Kam}. 

In section \ref{sec:2}, it is shown that 
on $\T^d$ T-duality group (originally 
for closed string theory\cite{GPR,MSch}) preserves 
the mass spectrum of open strings with both ends 
on the same D-brane. 
The transformation for the oscillator modes preserving the mass spectrum 
is also derived. 

In section \ref{sec:3}, we discuss about the T-duality transformation 
of D-branes from the viewpoint of the D-brane mass. 
It is shown that 
by the action of T-duality group D-branes and open strings 
with both ends on them are transformed together 
(to other D-branes and open strings with both ends on them). 

On two-tori, the T-duality group $SO(2,2 |\Z)$ decomposes to 
$SL(2,\Z)\times SL(2,\Z)$ and the arguments become very simple. 
We apply the above general arguments on two-tori and show that it 
agrees with some expected results in section \ref{sec:4}. 

The physical meaning of the above results, especially the relation 
to Morita equivalence is discussed in Conclusions and Discussions.

%

 \section{T-duality for open strings}
\label{sec:2}

Consider a bosonic open string on torus $\T^d$ in flat background 
$g$ and $B$ with its periodicity $2\pi$ 
for all $d$-direction. The action is  
\begin{equation}
 S=\int d\tau\int_0^{2\pi} d\sigma L
 =\ov{4\pi}\int d\tau\int_0^{2\pi} d\sigma 
 \Big[g_{\mu\nu}\partial_\alpha X^\mu\partial^\alpha X^\nu
     +\epsilon^{\alpha\beta} 
      B_{\mu\nu}\partial_\alpha X^\mu\partial_\beta X^\nu \Big]
\label{action}
\end{equation}
where $\mu, \nu$ run the Euclidean space direction $1,\cdots, d$,  
$\alpha, \beta$ are $\tau$ or $\sigma$ and its signature is Lorentzian : 
$(\tau, \sigma)=(+,-)$ .

The canonical momentum from the action (\ref{action}) is given by 
\begin{equation}
 P_\mu=\fdel{S}{X^\mu}=\ov{2\pi}
 \left(g_{\mu\nu}\Xd^\nu+B_{\mu\nu}\Xa^\nu\right)=\ov{2\pi}p_\mu+\cdots
 \label{mom}
\end{equation}
where $\Xd:=\fpart{X}{\tau}$, $\Xa:=\fpart{X}{\sigma}$, 
$p_\mu$ is the zero mode momentum, and $\cdots$ means 
the higher oscillator modes.
The Hamiltonian is 
\begin{equation}
 H=\int_0^{2\pi}d\sigma \big[P_\mu X^\mu-L\big]
  =\ov{4\pi}\int_0^{2\pi}d\sigma
 \big[\Xd^\mu g_{\mu\nu}\Xd^\nu+\Xa^\mu g_{\mu\nu}\Xa^\nu\big]\ .
 \label{H}
\end{equation}

Though in this section we will show the invariance of the open string 
mass spectrum under the T-duality including the oscillator parts, 
at first we concentrate on the zero mode parts of the open string. 
Then $X$ is represented as 
\begin{equation}
 X^\mu=x^\mu+f_\tau^\mu\tau+f_\sigma^\mu\sigma
 \label{zero}
\end{equation}
where $f_\tau^\mu$ and $f_\sigma^\mu$ are constant but linearly dependent 
because of boundary conditions for open strings. 

In the beginning, we consider the following standard boundary condition 
at $\sigma=0, 2\pi$  
\begin{equation}
 \begin{array}{rl}
 \mbox{Neumann b.c.}\ &: \quad g_{i\nu}\Xa^\nu+B_{i\nu}\Xd^\nu=0\\
 \mbox{Dirichlet b.c.}\ &: \quad \Xd^a=0 \ , 
 \end{array}
 \label{bc}
\end{equation}
where $\{i\}\cup\{a\}=\{1,\cdots,d\}$, $i$ is for Neumann b.c., and 
$a$ is for Dirichlet b.c. . As will be seen later, 
by T-duality transformation it transforms to other boundary conditions. 
Here we start with this standard one and derive the zero-mode 
open string mass systematically. 
Substituting eq.(\ref{zero}) into eq.(\ref{bc}) leads 
\begin{equation}
 g_{i\nu}f_\sigma^\nu+B_{i\nu}f_\tau^\nu=0\ ,\qquad f_\tau^a=0
 \label{bczero}
\end{equation}
and from eq.(\ref{mom}) with eq.(\ref{zero}) one gets 
$p_\mu=g_{\mu\nu}f_\tau^\nu+B_{\mu\nu}f_\sigma^\nu$. 
Moreover on tori, the zero mode momentum 
$p_\mu$ for Neumann b.c. direction and 
$f_\sigma^\mu$, which is the length of the open string over $2\pi$, 
for Dirichlet b.c. direction are quantized in integer, i.e. 
\begin{equation}
 g_{i\nu}f_\tau^\nu+B_{i\nu}f_\sigma^\nu = p_i \in \Z\ ,
 \qquad f_\sigma^a\in\Z \label{pmint}
\end{equation}
for each $i$ or $a$. 
Combining the boundary conditions (\ref{bczero}) 
and the integral conditions (\ref{pmint}) leads 
\begin{equation}
 \left[ \left(
 \begin{array}{cc}
  e & \\
    & e 
 \end{array}\right)
 \left(
 \begin{array}{cc}
  g & B\\
  B & g 
 \end{array}\right)
 +\left(
 \begin{array}{cc}
  \1-e & \\
    & \1-e 
 \end{array}\right)
 \right] 
  \left(
 \begin{array}{c}
  f_\tau \\
  f_\sigma 
 \end{array}\right)  
 =
 \left(
 \begin{array}{c}
  e\cdot p\\
  (\1-e)\cdot m 
 \end{array}\right)
 \label{M}
\end{equation}
where $p\in\Z^d$ is the momentum on $\T^d$, $m\in\Z^d$ is the winding number 
($\propto$ length)of the open string, 
and $e$ is a projection from $\{1,\cdots,d\}$ to Neumann b.c. directions 
$\{i\}$, i.e. $e\in \Mat(d,\Z)$
\footnote{$\Mat(d,\Z)$ means the set of $d\times d$ matrices with integer 
entries.}
 is a diagonal matrix with diagonal entries 
$1$ (for Neumann b.c.) or $0$ (for Dirichlet b.c.). 
Furthermore act 
\begin{math} T_e:=\bp e&\1-e\\ \1-e&e\ep\end{math}
on both sides of (\ref{M}) and we gets 
\begin{equation}
 M\bp f_\tau\\ f_\sigma \ep=\bp q \\ 0\ep\ ,\qquad 
 M=\bp eg & eB +(\1-e)\\ eB +(\1-e)&eg \ep \label{OSD}
\end{equation}
for $q=e\cdot p+(\1-e)\cdot m$. 
Hereafter let $q\in\Z^d$ be the degree of freedom of the open string 
zero-modes. This $M$ represents the open string with both ends 
on D$d'$-brane for $d'=\rank (e)$. 
Because $M$ decides the type of the open strings, 
we will call $M$ `open string data (OSD)' 
(and the form of $M$ is generalized later).   

Once OSD $M$ is given, the open strong zero-mode mass is obtained by 
substituting the OSD (\ref{OSD}) into the Hamiltonian(\ref{H}) 
\begin{equation}
 H_0=\ov{2}\bp q^t& 0\ep  \cM_o \bp q\\ 0 \ep  \ ,\qquad 
 \cM_o:=M^{t,-1} \bp g  & \0\\ \0 & g \ep M^{-1} \ . \label{HM}
\end{equation}

In particular case when $d'=d$ (all N b.c. ), 
OSD (\ref{OSD}) and its Hamiltonian (\ref{HM}) are 
\begin{equation}
 M=
 \bp
  g&B\\
  B&g
 \ep
\ ,\qquad 
 H_0=\ov{2}\bp q^t & 0 \ep \bp G^{-1} &  \\  & G^{-1} \ep \bp q\\ 0 \ep
 \label{aN}
\end{equation}
where $G:=g-Bg^{-1}B$ is the open string metric defined in \cite{SW}. 
This implies that an open string on a D$d$-brane
\footnote{Though noncommutativity or NC tori is not introduced 
in this paper, this D$d$-brane is in fact on a NC torus 
(with metric $G$), i.e. this open string theory is defined 
on a NC torus.} 
feels the open string metric. 
Conversely if $d'=0$ (all D b.c.), then 
\begin{equation}
 M=\bp {\bf 0}&{\bf 1}\\ {\bf 1}&{\bf 0} \ep\ ,\qquad 
 H_0=\ov{2}\bp q^t & 0\ep\bp g& \\ &g\ep\bp q\\ 0\ep\ .\label{aD}
\end{equation}
$B$ field does not affect the open string with its ends on the D$0$-brane, 
and the energy is proportional to its length
\footnote{This open string theory can also be defined 
on a NC torus. 
The torus has metric $g^{-1}$, which is related to the metric $G$ 
in eq.(\ref{aN}) by T-duality $T=T_{e=\0}$ 
defined below. The open string theories corresponding to eq.(\ref{aN}) are 
given by so-called Seiberg-Witten map (D$d$ $\raw$ NC D$d$) \cite{SW} and 
those corresponding to eq.(\ref{aD}) are given by compactifying 
matrix models (D$0$ $\raw$ NC D$d$) \cite{CDS,Ho,MZ,BMZ,PS,KO,CK,KS}. }. 

Here we consider the action of T-duality on $\T^d$\cite{GPR}. 
Let $E:=g+B\in \Mat(d,\R)$. 
The symmetric part (resp. antisymmetric part) of the matrix 
$E\in \Mat(d,\R)$ is $g$ (resp. $B$).
T-duality acts on the closed string background $E$ as 
\begin{equation}
 T(E)=\left(\cA(E)+\cB\right)\left(\cC(E)+\cD\right)^{-1} \label{Tt}
\end{equation}
where $T$ is an elements of $O(d,d |\Z)$ defined as 
\begin{equation}
 T^t\bp {\bf 0}&{\bf 1}\\ {\bf 1}&{\bf 0}\ep T=
  \bp {\bf 0}&{\bf 1}\\ {\bf 1}&{\bf 0}\ep\ ,\qquad
 T=\bp \cA &\cB \\ \cC & \cD \ep \ ,
 \qquad \cA, \cB, \cC, \cD \in \Mat(d,\Z)\ . \label{T}
\end{equation}
It is known that T-duality group $O(d, d | \Z)$ is generated by  
following three type of generators \cite{GPR,S} 
\begin{equation}
 T_e:= \begin{pmatrix}
              e& \1-e \\
              \1-e & e
             \end{pmatrix}
 \ ,\qquad 
 T_A= \begin{pmatrix}
           A      & {\bf 0} \\
           {\bf 0}& {A^t}^{-1}  
          \end{pmatrix}
 \ ,\qquad 
 T_N= \begin{pmatrix}
          \1 & N  \\
          \0 & \1 
      \end{pmatrix}
 \label{gen}
\end{equation}
for $A\in \mathit{GL}(d,\Z)$, $N^t=-N\in\Mat(d,\Z)$ and
$e\in\Mat(d,\Z)$ is a projection defined previously below eq.(\ref{M}).

We would like to find the transformation of OSD $M$ 
preserving the mass $H_0$ under the T-duality action (\ref{Tt}). 
In this paper we call such transformation as T-duality transformation 
of OSD $M$ (or open strings ). In order to find it, here we observe 
the action of these three generators. 

1.\ $T_{e'}$\quad 
$T_e$'s satisfy $T_{e'}T_e=T_{e''}$ for $e'':=e'e+(\1-e')(\1-e)$ 
where $e''\in\Mat(d,\Z)$ is also the projection. 
Let $M[E,T_e]$ be the OSD charactarized by $e$ with background $E=g+B$. 
Then it is natural to expect that the open string with OSD $M[E,T_e]$ 
is transformed to the one with OSD $M[T_{e'}(E),T_{e''}]$ by the action 
of T-duality $T_{e'}$. 
Actually it is. Let $\cM^{-1}_o[E,T_e]$ be the inverse of $\cM_o$ defined 
in eq.(\ref{H}) with $M=M[E,T_e]$. Then 
the direct calculation shows that 
\begin{equation}
 \cM^{-1}_o[E,T_e]=\cM^{-1}_o[T_{e'}(E),T_{e''}]\no
\end{equation}
is satisfied. (This equation will be shown also in eq. (\ref{proof}) 
where the arbitrary forms of OSD (\ref{GOSD}) 
which are compatible with the T-duality group are concerned. )
Thus one can see that the mass of the open string 
with OSD $M[E,T_e]$ (on the D$d'$-brane in background $E=g+B$) 
is equal to the mass of the another open string on the D$d''$-brane 
in background $T_{e'}(E)$. 
In particular if $e'=e$ then $e''=1$, which means 
that OSD $M'$ is transformed to the one 
of which the boundary condition is all-Neumann (\ref{aN}) 
in background $T_e(E)$, 
and in contrast in the case $e'=1-e$ then $e''=0$ and 
the boundary condition becomes all-Dirichlet (\ref{aD}) 
(in background $T_{(1-e)}(E)$). 
Moreover when $g={\bf 1}$ and $B=0$, 
the boundary condition part of these arguments 
reduce to those in \cite{DLP} and the physical picture is the same as 
that in \cite{T}. 

Anyway it was shown that the open string mass spectrum 
with OSD of the type $M[E,T_e]$ is invariant under the action of 
the subgroup $\{T_e'\}$.

2. $T_A$\quad Next consider the action of $T_A$ on the open string 
with OSD $M[E,T_e]$. Because $E$ transforms to $T_A(E)=AEA^t$, 
if $M[E,T_e]$ is transformed as
\begin{equation}
 M'= \left[\bp e A^{-1}& \\ &e A^{-1}\ep
 \bp T_A(g)&T_A(B)\\ T_A(B)&T_A(g)\ep \bp A^{t,-1}& \\ &A^{t,-1}\ep
 +\bp & 1-e\\ 1-e & \ep\right] \bp A^t& \\ &A^t\ep\ , \label{MA}
\end{equation}
then $\cM_o^{-1}$, or equivalently the mass $H_0$ is preserved.
Note that the $A^t$ acting from right in eq.(\ref{MA}) came from 
rewriting $g^{-1}= A^tT(g^{-1})A^{t,-1}$ in $\cM_o^{-1}$.     

Check the physical meaning of the transformed $M'$. 
\begin{math}
 M'\bp f'_\tau\\ f'_\sigma\ep=\bp q \\ 0\ep
\end{math} 
leads the following equations
\begin{equation*}
 \begin{split}
  eA^{-1} (T(g)f'_\tau +T(B)f'_\sigma)=e\cdot q\ ,&\quad
  e A^{-1} (T(B)f'_\tau +T(g)f'_\sigma)= 0 \\
  (1-e)A^t f'_\sigma=(1-e)A^t q\ ,&\quad
  (1-e)A^tf'_\tau=0\ .
 \end{split}
\end{equation*}
By comparing these equations with eqs.(\ref{bczero})(\ref{pmint}), 
one can see that the open string with OSD $M'$ is 
the one on the D$d'$-brane which winds nontrivially on $\T^d$ due to $A$ 
for $d'=\rank (e)$. We will write this $M'$ as $M[T_A(E),T_AT_e]$.

3. $T_N$ \quad Furthermore we act $T_N$ on the open string with OSD 
$M[T_A(E),T_AT_e]$. Here we rewrite $T_A(E)$ as $E'$. $T_N$ acts 
on $E'=g'+B'$ as $T_N(g')=g'$ and $T_N(B')=B'+N$. 
Because $T_N$ preserves $g'$, rewriting $M[E',T_AT_e]$ as 
\begin{equation}
 M''=\bp e A^{-1}& \\ &e A^{-1}\ep
 \bp T_N(g')&T_N(B')-N\\ T_N(B')-N&T_N(g')\ep
 +\bp & 1-e\\ 1-e & \ep \bp A^t& \\ &A^t\ep\ , \label{MN}
\end{equation}
and the mass of the open string is invariant. 
The meaning of this consequence is clear. $-N$ is the $U(1)$ 
constant curvature $F$ on $\T^d$. As is well-known, 
the curvature affects only on the direction of Neumann boundary condition, 
and the fact corresponds to the $eA^{-1}$ in eq.(\ref{MN}). 
The action of $T_N$ preserves the value of the pair $B+F$, and such a 
transformation is, on $\R^d$, often called as `$\Lambda$-symmetry'. 
However on $\T^d$, the elements of $F$ must be integer 
by the topological reason, and therefore the symmetry is discretized to 
be the group $\{T_N\}$. 

\vspace*{0.3cm}

From these three observations, it is natural to regard  
all the open strings in this system as those which are connected to 
the open string of all-Neumann 
b.c.(\ref{aN}) by any T-duality transformation.
Moreover extending these three example of OSD
(\ref{OSD})(\ref{MA})(\ref{MN}), 
the general form of the OSD which are given by acting $T$ on 
$M[T^{-1}(E),\1]$ (all-Neumann b.c.) can be expected to be the form 
\begin{equation}
  M[E,T]\bp f_\tau\\ f_\sigma \ep =\bp q \\ 0\ep\ ,\qquad 
  M[E,T]:=\bp \cAb g& \cAb B + \cBb\\ \cAb B + \cBb&\cAb g \ep
 \label{GOSD}
\end{equation}
for \begin{math}T^{-1}=: \bp \cAb&\cBb\\ \cCb&\cDb\ep
=\bp \cD^t&\cB^t\\ \cC^t&\cA^t\ep \end{math}
\footnote{The second equality follows from the definition of $O(d,d |\Z)$ 
(\ref{T}). }. 
Of course $\cM_o[E,T]$ and the mass $H_0$ are defined as the form in 
eq.(\ref{H}). 
As will be shown below, 
the mass spectrum given by this OSD is invariant under 
the T-duality transformation. In this sense, the form of OSD seems to be 
almost unique
\footnote{The boundary condition part of this OSD(\ref{GOSD}) can also 
be derived by applying the arguments in \cite{Kam} in this situation. }. 
We will show that this OSD is compatible with 
the T-duality transformation for D-branes in the next section 
and that on two-tori this derives expected results in the last section. 

In order to confirm that actually the mass spectrum for 
general open strings characterized by OSD(\ref{GOSD}) is compatible 
with the $O(d,d |\Z)$ T-duality action, we check that 
$\cM_o^{-1}$ is preserved under the action 
like as the above three examples. 
Define matrix \begin{math}
\cJ:=\ov{\sqrt{2}}\bp \1&\1 \\ \1&-\1\ep\in\Mat(2d,\Z)\end{math} 
which satisfies $\cJ^2=\1$ 
and $M[E, T]$ is rewritten as \begin{math}
M[E, T]=\cJ \bp \cAb E+\cBb & \\ &\cAb E^t-\cBb\ep\cJ\end{math}. 
Then one gets 
\begin{equation}
 \begin{split}
 \cM_o^{-1}[E,T]&=\cJ
             \bp \cAb E+\cBb & \\ &\cAb E^t-\cBb\ep
             \bp g^{-1}& \\ &g^{-1}\ep
             {\bp \cAb E+\cBb & \\ &\cAb E^t-\cBb\ep}^t \cJ\\
 &=\bp T^{-1}(E g^{-1} E^t)& \\
                & T^{-1}(E g^{-1} E^t)\ep
           =\bp T^{-1}(G) & \\ &T^{-1}(G) \ep \no
 \end{split}
\end{equation}
where 
\begin{math} T^{-1}(E)=(\cAb E+\cBb)(\cCb E+\cDb)^{-1}\ ,\ 
 T^{-1}(g^{-1})=(\cCb E+\cDb)g^{-1}(\cCb E+\cDb)^t
\end{math}. In the second step we used the identity 
$(\cAb E^t-\cBb)g^{-1}(\cAb E^t-\cBb)^t
=(\cAb E+\cBb)g^{-1}(\cAb E+\cBb)^t$. 

Thus, it has been shown that 
the mass of open strings with any OSD of the form 
(\ref{GOSD}) is equal to the mass of 
the open string with all-Neumann boundary condition 
(\ref{aN}) in background $T^{-1}(E)$ i.e. 
$\cM_o^{-1}[E,T]=\cM_o^{-1}[T^{-1}E,\1]$. 
Of course, this result means that any 
open string with OSD $M[E,T]$ (any $T$) 
translate to that with another OSD $M[T'(E),T'T]$ 
in another T-dual background $T'(E)$ and the mass is preserved
\begin{equation}
 \cM_o[E,T]=\cM_o[T'(E),T'T]\ . \label{proof}
\end{equation}
Mention that a certain subset of T-duality group acts trivially 
on open strings, because the open strings have 
half degree of freedom compared with the closed strings. 

In the last of this section, we will derive the T-duality transformation 
for the oscillator parts. The mode expansion of $X$ is 
\begin{equation*}
 X^\mu=x^\mu+f_\tau^\mu\tau+f_\sigma^\mu\sigma+\sum_{n\ne 0}
 \frac{2 e^{\frac{-i n\tau}{2}}}{n}
 (i a_{(n)}\cos(\frac{n\sigma}{2})+b_{(n)}\sin(\frac{n\sigma}{2}))\ , 
\end{equation*}
and substituting this into the Hamiltonian (\ref{H}) leads 
\begin{equation}
 H=\half \bp q^t&0\ep\cM_o \bp q\\ 0 \ep+\half\sum_{n\ne 0}
 \bp a^t_{(n)}&b^t_{(n)}\ep \bp g^{-1}& \\ &g^{-1}\ep 
 \bp a_{(-n)}\\b_{(-n)}\ep\ . \no 
\end{equation}
For open strings, the $a_{(n)}$ and $b_{(n)}$ are linearly dependent 
because of the constraint from boundary conditions. 
The general form of boundary conditions are 
\begin{equation*}
 (\cAb B+\cBb)\Xd+\cAb g\Xa \Big|_{\sigma=0,2\pi}=0\ ,\qquad 
 T\in O(d,d |\Z) 
\end{equation*}
or for each mode $n$, $(\cAb B+\cBb)a_{(n)}+\cAb g b_{(n)}= 0$. In the 
same spirits of the OSD(\ref{GOSD}), let us define $q_{(n)}$ as 
\begin{equation}
 M[E,T]\bp a_{(n)}\\ b_{(n)}\ep= \bp q_{(n)}\\  0\ep\ ,\qquad  
 M[E,T]:=\bp \cAb g& \cAb B + \cBb\\ \cAb B + \cBb&\cAb g \ep\ ,
 \label{OSDosc}
\end{equation}
and the Hamiltonian can be represented as the following form
\begin{equation}
 H=\half \bp q^t&0\ep\cM_o \bp q\\ 0 \ep+\half\sum_{n\ne 0}
 \bp q^t_{(n)}&  0\ep \cM_o
 \bp q_{(-n)}\\  0\ep\ . \no  
\end{equation}
As was already shown, when background $E$ transforms to $T'(E)$, 
$\cM_o$ is invariant if OSD $M[E,T]$ changes to $M[T'(E),T'T]$.
Therefore the mass is invariant if each $q_{(n)}$ is preserved under 
the transformation $T'$. Thus the transformations 
for $a_{(n)}$ and $b_{(n)}$ are given by 
acting $M[E, T]^{-1}$ on both sides of eq.(\ref{OSDosc}) similarly for the 
zero modes $f_\tau, f_\sigma$ in eq.(\ref{GOSD}).

 \section{T-duality for D-branes}
\label{sec:3}

In this section, we discuss the T-duality transformation for D-branes 
and show that the transformation agrees with 
the consequence in the previous section. 

T-duality group acts on $g_s$ as 
\begin{equation*}
  T(g_s)=g_s\det{(\cC E+\cD)^{-\half}}\ .
\end{equation*}
The mass of the D$d$-brane is the constant term of the DBI-action 
\begin{equation*}
 \M_{D}[E,{\bf 1}]=\ov{g_s\sqrt{\alpha'}}\sqrt{\det E}\ , 
\end{equation*}
and which can be rewritten with the variable in background $T(E)$ as 
\begin{equation}
 \M_{D}[E,{\bf 1}]=\ov{T(g_s)\sqrt{\alpha'}}\sqrt{\det (\cAb T(E)+\cBb)}
 =:\M_{D}[T(E), T]
 \ ,\qquad T^{-1}=:\bp \cAb &\cBb \\ \cCb &\cDb \ep\ .
 \label{Dtf}
\end{equation}
This identities is considered as the T-duality transformation for the 
D-brane mass corresponding to that for the open string 
(\ref{GOSD})(\ref{OSDosc})
\footnote{This D-brane transformation is in fact compatible with that 
on NC tori. 
In two-tori case these arguments are given in \cite{KMT} and applied to 
the T-duality for NC solitons on two-tori. 
In the $|B/g|\raw\infty$ limit, the arguments reduce to those 
in \cite{BKMT}.}. 
The right hand side of eq.(\ref{Dtf}) 
is regarded as the mass of some D-brane bound state in background $T(E)$ 
(and which implies that the D-brane mass spectrum is invariant 
under the T-duality). 

In order to clarify the states represented by the 
right hand side in eq.(\ref{Dtf}), 
here let us consider as $T$ in eq.(\ref{Dtf}) the three type 
of generators(\ref{gen}) or some compositions of those 
as discussed previously. 

\vspace*{0.3cm}

1.\ $T_e$\quad Take $T$ as $T_e$ and 
the right hand side in (\ref{Dtf}) becomes
\begin{equation}
 \ov{g_s\sqrt{\alpha'}}\sqrt{\det (e E+(1-e))}
 =\ov{g_s\sqrt{\alpha'}}\sqrt{{\det}_e(E)}=\M_{D}[T(E),T] \label{DTe}
\end{equation}
where $\det_e (E)$ means the determinant for the submatrix of $E$ 
restricted on the elements for the Neumann b.c. direction.
This is exactly the mass of the D$d'$-brane in background $T_e(E)$ 
for $d'=\rank (e)=\sharp \{i\}$. 
On this D$d'$-brane the open strings with OSD $M[T_e(E),T_e]$ live. 

2.\ $T_A$\quad Here replace the above $T_e(E), T_e(g_s)$ with $E, g_s$ 
and consider acting $T_A$ on eq.(\ref{DTe}) in parallel with the 
previous arguments in section \ref{sec:2}. 
 The mass of the single D$d'$-brane in background $E$ is 
\begin{equation}
 \ov{T_A(g_s)\sqrt{\alpha'}}\sqrt{\det(eA^{-1}T_A(E)+(1-e)A^t)}
 =\M_{D}[T_A(E),T_AT_e]
 \label{DTA}
\end{equation}
in background $T_A(E)$
\footnote{The mass corresponds to the $\M_{D}[T(E),T]$ 
in eq.(\ref{Dtf}) 
with replacing $E$ with $T_e^{-1}(E)$ and $T=T_AT_e$ . }
. Because of 
\begin{equation}
 \begin{split}
 \sqrt{\det(eA^{-1}T_A(E)+(1-e)A^t)}
 &=\sqrt{\det(eA^{-1}(T_A(E))A^{t,-1}+(1-e))}\\
 &=\sqrt{{\det}_e(A^{-1}T_A(E)A^{t,-1})}
 \end{split}\no
\end{equation}
and comparing with the OSD(\ref{MA}), 
one can see that this is the mass of the D$d'$-brane which 
winds nontrivially on the torus with background $T_A(E)$ 
due to the transformation of $A$.

3.\ $T_N$\quad Further acting $T_N$ on eq.(\ref{DTA}) yields 
\begin{equation}
 \ov{g_s'\sqrt{\alpha'}}\sqrt{\det(eA^{-1}(T_N(E')-N)+(1-e)A^t)}
 =\M_{D}[T_N(E'),T_N T_A T_e] \no
\end{equation}
for $E'=T_A(E)$ and $g_s'=T_A(g_s)\ (=T_N(g'_s))$. This is also the 
expected results and is consistent with the corresponding OSD(\ref{MN}). 
On the D$d'$-brane twisted on $\T^d$ by $A$, the line bundle over it 
is also twisted and 
the open strings with OSD $M''=M[T_N(E'),T_NT_AT_e]$ are on it. 

\vspace*{0.2cm}

One has seen that both the open string with generic OSD and 
the D-branes on which the open string ends transform together 
consistently in the examples where the number of 
the D-brane of the highest dimension is one. 

In general, the D-brane of the highest dimension can be more than one. 
We will see such an example in the $\T^2$ case.

 \section{T-duality on two-tori}
\label{sec:4}

$SO(d,d |\Z)$, which is given by restricting the rank of $\1-e$ in 
$T_e\in O(d,d |\Z)$ to even rank, is known as the group which arises in 
the issue of Morita equivalence on NC $\T^d$. 
Furthermore, when the dimension of the torus is two, 
it is well-known that it can be decomposed 
as $SO(2,2 |\Z)\simeq SL(2,\Z)\times SL(2,\Z)$. 
One of two $SL(2,\Z)$ groups is the 
modular transformation of the target space $\T^2$. It corresponds to 
$T_A\ (A\in SL(2,\Z))$ in $SO(2,2 |\Z)$ and 
preserves $g_s$ and $\sqrt{\det(E)}$. 
The other $SL(2,\Z)$ is discussed as the group generating 
Morita equivalent NC $\T^2$ \cite{CDS,PS,KS} 
and here at first we discuss the action of this part. 
This $SL(2,\Z)$ transformation can be embedded into 
$SO(2,2 |\Z)$ as follows
\begin{equation}
 \ba{ccc}
  SL(2,\Z) & \hookrightarrow & SO(2,2 |\Z)\\
  t^+_2:=\bp a&b\\c&d \ep &\mapsto  
  & \bp a{\bf 1}&b \J\\ c(-\J)& d{\bf 1}\ep=:t^+ 
 \ea \label{I}
\end{equation}
for 
\begin{math}\J:=\left(\begin{smallmatrix}
0&1\\-1&0\end{smallmatrix}\right)\in \Mat(2,\Z)\end{math}
\footnote{This $SL(2,\Z)\subset SO(2,2 |\Z)$ is essentially generated 
by $\{T_e\}$ and $\{T_N\}$ in the following sense. 
When $d=2$, the subgroup $\{T_e\}$ ( resp. $\{T_N\}$) is generated by 
\begin{math}\l( \bps \0& \1\\ \1& \0\eps \r)\end{math} 
( resp. \begin{math} \l( \bps \1&\J\\ \0&\1\eps \r)\end{math}). 
On the other hand, it is known that $SL(2,\Z)$ is generated by 
\begin{math}\l( \bps 0&1\\ -1&0\eps \r) \end{math} and 
\begin{math}\l( \bps 1&1\\ 0&1 \eps \r) \end{math}, which are embedded 
into $SO(2,2 |\Z)$ as 
\begin{math}\l( \bps 0&\J\\ \J&0\eps \r)\end{math} and 
\begin{math}\l( \bps {\bf 1}&\J\\ 0&{\bf 1}\eps \r)\end{math} 
, respectively. 
\begin{math}\l( \bps \0& \1\\ \1& \0\eps \r)\end{math} and 
\begin{math}\l( \bps 0&\J\\ \J&0\eps \r)\end{math} are related by 
the modification of 
\begin{math} T_{A=\J}:=\l( \bps \J &0\\ 0&\J\eps \r) \in\{T_A\}\end{math} 
as \begin{math}\l( \bps \0&\J\\ \J&\0\eps \r)=T_{A=\J}
\l( \bps \0& \1\\ \1& \0\eps \r)\end{math}. 

Note that any element of $SO(2,2|\Z)$ can be written 
as the form $t^-t^+$ for some $t^-\in \{T_A\}$ and 
\begin{math}t^+\in \{\l( \bps 0&\J\\ \J&0\eps \r), 
\l( \bps {\bf 1}&\J\\ 0&{\bf 1}\eps \r)\}\end{math}. }. 

The general OSD (\ref{GOSD}) 
and D-brane mass (\ref{Dtf}) are simplified as 
\begin{equation}
 \begin{split}
 &\bp d\cdot g& d\cdot B -b\J\\ d\cdot B -b\J&d\cdot g \ep
 \bp f_\tau\\ f_\sigma\ep
 =\bp q \\ 0 \ep \ ,\qquad q\in\Z^2\ ,\\
 &\M_{D}[E, t^+]=\ov{g_s\sqrt{\alpha'}}\sqrt{\det(d E -b\J)}\ .
 \end{split}\label{T2}
\end{equation}
This is the system of the 
$(d, -b)$ D2-D0 bound state in background $E$. 
When $d\ge 1$, the system is described by $\rank\ d$ gauge theory with 
the constant curvature.

The OSD generated by all the elements of $SO(2,2|\Z)$ are 
given by acting $t^-\in\{T_A | A\in SL(2,\Z)\}$ on eq.(\ref{T2}). 

One more interesting example is the system derived by acting 
$T_e$ with $\rank (e)=1$ on the above D$2$-D$0$ system. 
The system is transformed to the system of 
only a single D$1$-brane, and the arguments reduce to those discussed in 
\cite{AAS}. Let us discuss the connection finally. 

The OSD for this system is $M[E, T_e t^-t^+]$ for   
\begin{math}e= \l(\bps 1&0\\0&0\eps\r)\end{math} (fixed),  
$t^-=T_A$ with any 
\begin{math}A= \l(\bps p&q\\r&s\eps\r)\in SL(2,\Z)\end{math} and any 
$t^+$ of \begin{math}t^+_2=
\l(\bps a&b\\c&d\eps\r)\in SL(2,\Z)\end{math} in eq.(\ref{I}). 
Explicitly, with a little calculation, 
\begin{equation*}
 M[E, T_e t^-t^+]=
 \bp A^{-1}e (t^+)^{-1}g & A^{-1}e (t^+)^{-1}B+A^{-1}(1-e)(t^+)^t\\
     A^{-1}e (t^+)^{-1}B+A^{-1}(1-e)(t^+)^t & A^{-1}e (t^+)^{-1}g \ep\ .
\end{equation*}
Acting $A$ in both sides of 
\begin{math} M[E, T_e t^-t^+]\bp f_\tau\\ f_\sigma\ep=\bp q\\0\ep\end{math}
 leads 
\begin{equation*}
  M[E, t^+T_e]\bp f_\tau\\f_\sigma\ep=\bp A\cdot q\\ 0\ep\ .
\end{equation*}
This shows that the number of the D$1$-brane is exactly one. Moreover 
one can read that 
$A\in SL(2,\Z)$ acts as the automorphism for the zero-modes 
of the open strings and that 
$t^+$ twists the D$1$-brane on $\T^2$, as argued in \cite{AAS}. 
The previous general arguments (\ref{GOSD})(\ref{Dtf}) guarantee 
that the mass spectrum for the open strings and D$1$-brane are 
preserved under the two $SL(2,\Z)$. 

From the viewpoint of $K$-theory, the D$2$-D$0$ bound states in the 
former situation correspond to the elements of $K^0(\T^2)$, whereas 
the latter situation corresponds to $K^1(\T^2)$. 
It is interesting that $t^+$, which is the automorphism on $K^0(\T^2)$, 
is transformed to that on $K^1(\T^2)$ by $T_e$.

 \section{Conclusions and Discussions}
\label{CD}

We studied the T-duality group for open string theory on $\T^d$ with 
flat but general background $E=g+B$. 
We constructed the generic boundary conditions which are expected from 
the existence of $O(d,d|\Z)$ T-duality group, and 
showed that the mass spectrum of the open strings with 
those boundary conditions is invariant under the T-duality group. 
Furthermore, by discussing the D-brane mass spectrum which is 
invariant under the T-duality, we derived the T-duality transformation 
for D-branes. They should be the boundary of the open strings, 
and showed that actually the D-branes and the open strings on them 
transform together consistently.

Physically, from the viewpoints of field theories on the target space 
$\T^d$, the open string mass spectrum 
\begin{math} \half \l(\bps q^t & 0\eps\r)\cM_o\l(\bps q\\ 0\eps\r)
\end{math} corresponds to 
the kinetic term for field theories 
on NC $\T^d$. 
The open string zero-modes $q$ are, even if the 
degree of freedom are the length of open strings on the 
commutative $\T^d$, translated into the momentum of the fields on the 
NC $\T^d$. This picture seems to give us the realization 
of the Morita equivalence from the open string physics
\footnote{Precisely, when connecting the T-duality transformation 
on commutative $\T^d$ with that on NC $\T^d$ with finite 
$|B/g|$, we have to introduce background $\Phi$ 
on NC $\T^d$ \cite{S,SW,PS,KS,KMT}. }. 
The mass of the D-branes on $\T^d$ are also translated as 
that on the NC $\T^d$, which is the constant (leading) term 
of the DBI-action.

Precisely, when saying Morita equivalence, 
we have to study the noncommutativity, which is not discussed 
explicitly in the present paper. 
Because we now get the explicit form of 
the T-duality covariant open string modes, 
it is one approach to the issue along this paper 
to study the interaction terms 
for NC field theory similarly in \cite{KO,CK}. 

It is also interesting to comment about the Chan-Paton degree of freedom. 
T-duality is originally the duality for closed string theory. 
When applying this T-duality to open string theory, the gauge bundles 
emerge. 
They are generally twisted and the rank are generally greater than one, 
though only the $U(1)$ parts of the gauge group is concerned. 
This relation between closed string theory and open string theory is 
intriguing. Moreover, 
the T-duality for such pairs of submanifolds in $\T^d$ 
and the $U(1)$-bundle discussed in the present paper 
may be applied to other compactified manifolds.


\begin{center}
\noindent{\large \textbf{Acknowledgments}}
\end{center}

I am very grateful to A.~Kato for helpful discussions and advice.
I would also like to thank M.~Kato, T.~Takayanagi, and S.~Tamura 
for valuable discussions.
The author is  supported by JSPS Research Fellowships for Young
Scientists.


\end{document}